\documentclass[twocolumn]{article}

\usepackage{arxiv}

\usepackage[utf8]{inputenc} 
\usepackage[T1]{fontenc}    
\usepackage{hyperref}       
\usepackage{url}            
\usepackage{booktabs}       
\usepackage{amsfonts}       
\usepackage{nicefrac}       
\usepackage{microtype}      
\usepackage{lipsum}
\usepackage{graphicx}
 \usepackage{amsmath}
\usepackage{amssymb}

\usepackage{tipa}
\graphicspath{{imgs/}}
\title{Analytic description of anomalous diffusion in heterogeneous environments: Fokker-Planck equation without fractional derivatives}

\author{
 Ilia Kalashnikov \\
Keldysh Institute of Applied Mathematics, \\
4 Miusskaya sq., Moscow, 125047, Russia\\
  \texttt{kalasxel@gmail.com} \\
   \And
 Polina Likhomanova \\
National Research Center ''Kurchatov Institute'', \\
Akademika Kurchatova sq.1, Moscow, 123182, Russia.\\
  \texttt{likhomanovapa@gmail.com} \\
}

\begin{document}
	
	\twocolumn[{
	
\maketitle
\begin{abstract}
We present a model of diffusion in heterogeneous environment, which qualitatively reflects the transport properties of a polymeric membrane with carbon nanotubes. 
We derived Fokker-Planck equation from system of stochastic equations, responding diffusion regime in polymer and ballistic regime in nanotube areas. We show how probability density function is changed in presence of nanotube areas. Hereupon
nonlinear time dependence of mean square displacement is observed, that indicates anomalous diffusion regime. Thus, the model explains mechanism of appearance of anomalous diffusion for diffusion-ballistic regime. Our approach does not suppose any type of distributions and using fractional differentiate apparatus.
\end{abstract}
 \smallskip \smallskip \smallskip \smallskip\smallskip\smallskip\smallskip
	}]

\section{Introduction}

In the recent years the problem of diffusion in heterogeneous environments becomes more and more relevant. It is connected with expanding area of its application: biomembranes, cellular biology, geological researches, nanomaterials  etc. \cite{Javanainen2013AnomalousAN, H_fling_2013, Baum2014RetrievingTI, Matveev, doi:10.1002/aic.10998}. In particular, such problem arises for gaseous diffusion in polymeric membranes with carbon fillers, as example, carbon nanotubes (CNTs) \cite{Grekhov2013,BAKHTIN2015166,Kim2007HighPD,KIM2007147}. There are three ways of gas transport through the mixed matrix membranes: through dense layer of polymer matrix, highly selective carbon nanotubes and non-selective voids between the matrix and sieve particles \cite{Kusworo2012TheUO}. The last two can be considered as traps in which a particle can get stuck or vice versa pass the area without stopping and without changing its direction of movement \cite{677868145,Zelenyi:2004}. Thus, due to the addition of fillers complex diffusion regime is realized. It includes two factors: diffusion in polymer matrix and regime connected with moving in CNT areas. This may lead to a decreasing or increasing of transport characteristics.

For mixed matrix membranes based on poly(trimethylvinylsilane) and polysulfone, containing carbon nanotubes (CNT) by experimental methods was shown increasing of transport characteristics (permeability and diffusion coefficient) for tested gases $N_2, O_2, CH_4, C_3 H_8, CO_2$ \cite{Grekhov2013,BAKHTIN2015166,KIM2007147}. Both characteristics increase with increasing weight fraction of CNT. In \cite{Kim2007HighPD} it was assumed that carbon nanotubes provide high diffusivity tunnels in CNT regions. At the same time, in \cite{PhysRevLett.89.185901,Verweij2007FastMT} high diffusive transport rates in CNTs for light gases and possibility of transition from Fickian diffusion to ballistic regime \cite{C4CP03881A} were established. Thereby we may suppose that these tunnels allow gas particle to freely pass this region, in contrast to the polymer area. Such passages are associated with so-called Levy flights.

Often, such diffusion processes are characterized by a nonlinear time dependence of mean square displacement (MSD) and they are described by the expression $\langle x^2 \rangle \sim t^{\mu}$, where $\mu$ is exponent, characterizing deviation from normal diffusion regime. These processes are called anomalous diffusion.

Usually anomalous diffusion is attempted to be described using the apparatus of fractional differential equations, in particular, the fractional Fokker-Planck equation \cite{METZLER20001,Uchaikin_2003,SibUch09,Chukbar1,Chukbar2}. Fractional differential equations are directly related to the stable distributions of the particle on jump length and waiting time that underlie their derivation. However, physics reasons for {\it a priori} assumption for such type of distributions are not always clear.

For overcoming these ambiguities we suggest a model without assumptions about distributions type. In this model transfer regime in polymer matrix is considered as normal diffusion \cite{bekman2016matematika17023865}, and to describe the motion of a gas particle near to CNT we assume that a ballistic mode can be chosen \cite{MANTZALIS2014244,doi:10.1063/1.3532083}. Such a choice, as noted above, is associated with free passing of particle in CNT area. In case of ballistic regime a particle save its velocity during passage this area both in absolute value and in direction.

In our earlier work \cite{Likhomanova_2018} a similar model was realized by numerical simulation using continuum time random walk \cite{ABDELREHIM2008274} and Monte-Carlo method for 2D case. Using mechanism of particle motion described above the stable distributions for the particle displacements was obtained. Thus we showed that such mechanism of particle motion in polymer-CNT systems leads to anomalous diffusion law.

In the present study the same mechanism was investigated with analytic methods for 1D case. For this from stochastic equations for a particle moving in heterogeneous media, which respond diffusion-ballistic regime, Fokker-Planck equation was derived. Obtained solution of Fokker-Planck equation shows how CNT areas affect on probability density function (PDF) of particle displacement. The advantage of this approach is the rejection of using of the mathematical apparatus associated with fractional differential derivatives. It simplifies description of transport processes in heterogeneous environments and clarifies the occurrence of the anomalous diffusion regime.

\section{Equations for heterogeneous diffusion}

\begin{figure}[h]
	\center{\includegraphics[width=1\linewidth]{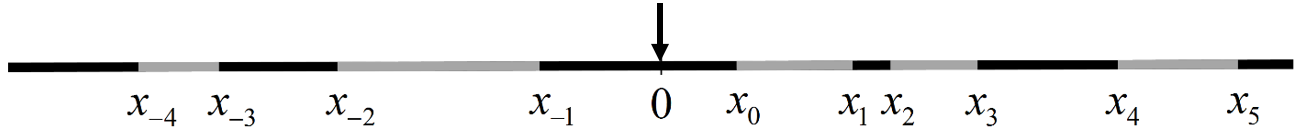}}
	\caption{1D heterogeneous system with alternating CNT's and polymer areas. CNT -- grey, polymer -- black. }
	\label{fig:fig1}
\end{figure}

Recently, multi-zone lattice random walk model was presented in \cite{Nazarenko_Blavatska_2018}. In the model various diffusion coefficients correspond to every zone. Our model is also divided on zones. However, particle motion is arranged otherwise. We considered combined diffusion process as random walk of one particle in 1D inhomogeneous continuous environment of infinite size, consisting of alternating areas of CNT and polymer. 
To construct this environment we divided the line with random located dots (see Fig. \ref{fig:fig1}). These dots respond to the boundaries of CNT's and polymer areas. We denoted these dots as $ \left\{ {x_0,x_1,x_2...} \right\}$ for $x>0$ and $ \left\{ {x_{-1},x_{-2}...} \right\}$ for $x < 0$. For definiteness we supposed that $x=0$ is belongs to the polymer area. Thus, next segment is CNT etc.

At the initial time the particle starts moving at $x=0$. The particle movement in such system is organized as follows: if the particle is in the polymer region, normal diffusion occurs, and in the regions corresponding to CNTs, the ballistic regime is realizes. The speed with which the particle moves in the CNTs is determined by the speed with which the particle flew into the CNTs. 

Now, random process, which is described this combined moving regime for presented 1D system may be written via system of stochastic equations:
\begin{align}
	\begin{split}
		&dx=vdt,  \\
		&dv=\chi(x)(-\gamma v dt +\sigma \delta W),
		\label{STCH}
	\end{split}
\end{align}
where $v$ -- particle velocity; $\gamma$ -- attenuation coefficient, inversely proportional to the characteristic decay time; $\chi(x)$ -- function, corresponding to CNT distribution on the line;
$W$ --  Wiener process and $\sigma$ is its dispersion.

Since transition from the stochastic equations (\ref{STCH}) to probability density equation requires smooth function, to avoid discontinuity of the function $\chi(x)$ we define it as:
\begin{align}
	\begin{split}
		&\chi(x) = 1-\sum_{k}\left(\theta(x-x_{2k})-\theta(x-x_{2k+1})\right),  \\
		&\theta(x) = \frac 1 2 \left(1+\tanh(mx)\right). 
		\label{chiDef}
	\end{split}
\end{align}
Then $\chi(x)$ is smooth function and for large value of $m$ it tends to $1$ for polymer regions and to $0$ for CNTs areas. But at the same time $\chi(x)$ is nowhere equals to zero.

Then probability density equation for heterogeneous system can be obtained from (\ref{STCH}):
\begin{equation}
\frac{\partial P}{\partial t} +v \frac{\partial P}{\partial x} -\gamma \chi \frac{\partial }{\partial v}vP -\frac {1} {2} \sigma^2 \chi^2 \frac{\partial^2 P}{\partial v^2} = 0.
\label{FP}
\end{equation}
If $\chi\equiv 1$, (\ref{FP}) is the so-called Klein–Kramers equation. Earlier, solution of this equation on infinite straight was obtained in \cite{Standish1990OnVQ}. Further we will call (\ref{FP}) for $\chi\not\equiv 1$ as modified Klein–Kramers equation.

For (\ref{FP}) we have the following initial and boundary conditions:
\begin{align}
	\begin{split}
		& P(t=0)=\delta(x)\delta(v-v_0),  \\
		& P(v\rightarrow \pm\infty)\rightarrow 0, \\
		& P(x\rightarrow \pm\infty)\rightarrow 0.
	\end{split}
	\label{FPoutBo1}
\end{align}

Now, since we not interested in distribution on $v$ and considered diffusion on large time scale, we went over from modified Klein-Kramers to Fokker-Planck equation. 

\section{From modified Klein-Kramers to Fokker-Planck equation}

According to (\ref{chiDef}) $\chi(x) \neq 0 $ for any $x$ and for enough large number of $m$ we have $\chi^2 \approxeq \chi$, which will be used hereinafter. Following \cite{article} let's get rid of the variable $v$, moving to long-time limit $t\gg\gamma^{-1}$. For this let's rewrite (\ref{FP}) in the following form:
\begin{equation}
\gamma^{-1} \left( \frac{\partial}{\partial t} +v \frac{\partial }{\partial x} \right) P =  \chi \frac{\partial }{\partial v} \left( v +\frac {\sigma^2} {2 \gamma} \frac{\partial }{\partial v} \right) P.
\label{FPexp}
\end{equation}
If we expand $P$ in degrees of $\gamma$:
\begin{equation}
P=P_0 + \gamma^{-1} P_1 + \gamma^{-2} P_2 +\mathcal{O}(\gamma^{-3}),
\end{equation}
and identify terms of the same order then we find the following equation for zero order:
\begin{equation}
\chi \left( \frac{\partial }{\partial v} v P_0 +\frac {\sigma^2} {2 \gamma} \frac{\partial^2 P_0 }{\partial v^2} \right) = 0,
\label{FPexp0}
\end{equation}
whence we get a solution
\begin{equation}
P_0=\phi(x,t) \exp \left( - \frac{\gamma}{\sigma ^2} v^2 \right),
\label{exp1}
\end{equation}
where $\phi$ is an arbitrary function.

The first order equation has the form:
\begin{equation}
\frac{\partial P_0}{\partial t} +v \frac{\partial P_0}{\partial x} =  \chi  \left(\frac{\partial }{\partial v} v P_1 + \frac {\sigma^2} {2 \gamma} \frac{\partial^2 P_1 }{\partial v^2} \right).
\label{exp2}
\end{equation}
Let's substitute (\ref{exp1}) into (\ref{exp2}):
\begin{equation}
( \dot{\phi} + v \phi ' ) \exp \left( - \frac{\gamma}{\sigma ^2} v^2 \right) = \chi \frac{\partial }{\partial v}  \left( v P_1 +\frac {\sigma^2} {2 \gamma} \frac{\partial P_1 }{\partial v} \right),
\end{equation}
where the dot and prime denote $t$ and $x$ derivatives respectively. By integrating both sides over $v$ in infinite limits we obtain the first solubility condition:
\begin{equation}
\dot{\phi} = 0.
\label{soli1}
\end{equation}
Then the first order correction can now be written as
\begin{equation}
P_1=\psi(x,t) \exp \left( - \frac{\gamma}{\sigma ^2} v^2 \right) - \phi'\frac{v}{\chi} \exp \left( - \frac{\gamma}{\sigma ^2} v^2 \right),
\label{FPexp2.0}
\end{equation}
with an arbitrary $\psi$.

Equation for the second order is following:
\begin{equation}
\frac{\partial P_1}{\partial t} +v \frac{\partial P_1}{\partial x} =  \chi  \left(\frac{\partial }{\partial v} v P_2 + \frac {\sigma^2} {2 \gamma} \frac{\partial^2 P_2 }{\partial v^2} \right). 
\label{FPexp2}
\end{equation}
Substituting (\ref{FPexp2.0}) into (\ref{FPexp2}) we have:
\begin{align}
	\begin{split}
		\frac{\partial P_1}{\partial t} +v \frac{\partial P_1}{\partial x} &=\left (\dot{\psi} + v \psi ' - \frac{v}{\chi} \dot{\phi}' - \right. \\
		&\left.- \frac{v^2}{\chi} \left( \phi '' - \frac{\chi '}{\chi}\phi ' \right) \right ) \exp \left( - \frac{\gamma}{\sigma ^2} v^2 \right). 
		\label{FPexp2.1}
	\end{split}
\end{align}
Integrating (\ref{FPexp2.1}) over $v$ in from $-\infty$ to $\infty$ we can get the second solubility condition:
\begin{eqnarray}
\dot{\psi} - \frac{\sigma ^2}{2 \gamma \chi} \left(\phi'' - \frac{\chi'}{\chi}\phi'\right)  =0. 
\label{FPsys}
\end{eqnarray}

Given (\ref{exp1})(\ref{FPexp2.0}) relations, we have
\begin{equation}
P=\left( \phi + \frac{\psi}{\gamma} -\frac{\phi '}{\chi \gamma} v \right) \exp \left( - \frac{\gamma}{\sigma ^2} v^2 \right) + \mathcal{O}(\gamma^{-2}).
\end{equation}
Since the probability density $\rho(x,t)$ reads
\begin{equation}
\rho (x,t) = \int_{-\infty}^{\infty} P(x,v,t) dv = \sqrt{\frac{\pi \sigma^2}{\gamma}} \left(\phi +\frac{\psi}{\gamma}\right),
\label{rhoT}
\end{equation}
then taking into account (\ref{soli1}) (\ref{rhoT}) second solubility condition (\ref{FPsys}) leads to Fokker-Planck equation:
\begin{equation}
\frac{\partial \rho}{\partial t} = D \frac{\partial}{\partial x} \frac{1}{\chi} \frac{\partial \rho}{\partial x},
\label{FPresult}
\end{equation}
where $D=\sigma^2/2\gamma^2$  - usual coefficient of diffusion in polymer. 

Note that equation (\ref{FPresult}) can be obtained using the technique of multiple time-scales presented in \cite{article}.

\section{Probability density function for heterogeneous environment}

\begin{figure}
	\begin{minipage}[h]{0.49\linewidth}
		\center{\includegraphics[width=1\linewidth]{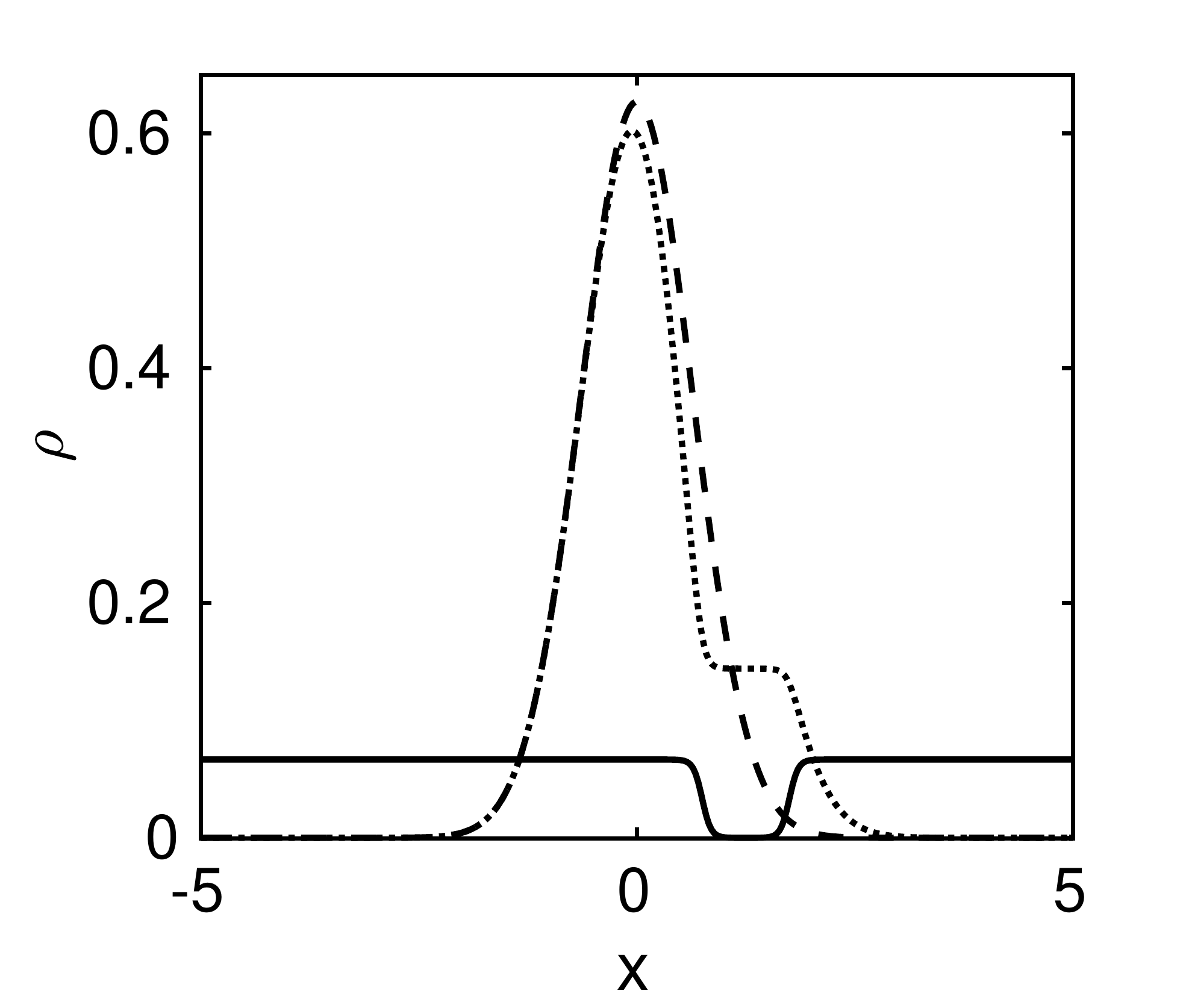}} a) \\
	\end{minipage}
	\hfill
	\begin{minipage}[h]{0.49\linewidth}
		\center{\includegraphics[width=1\linewidth]{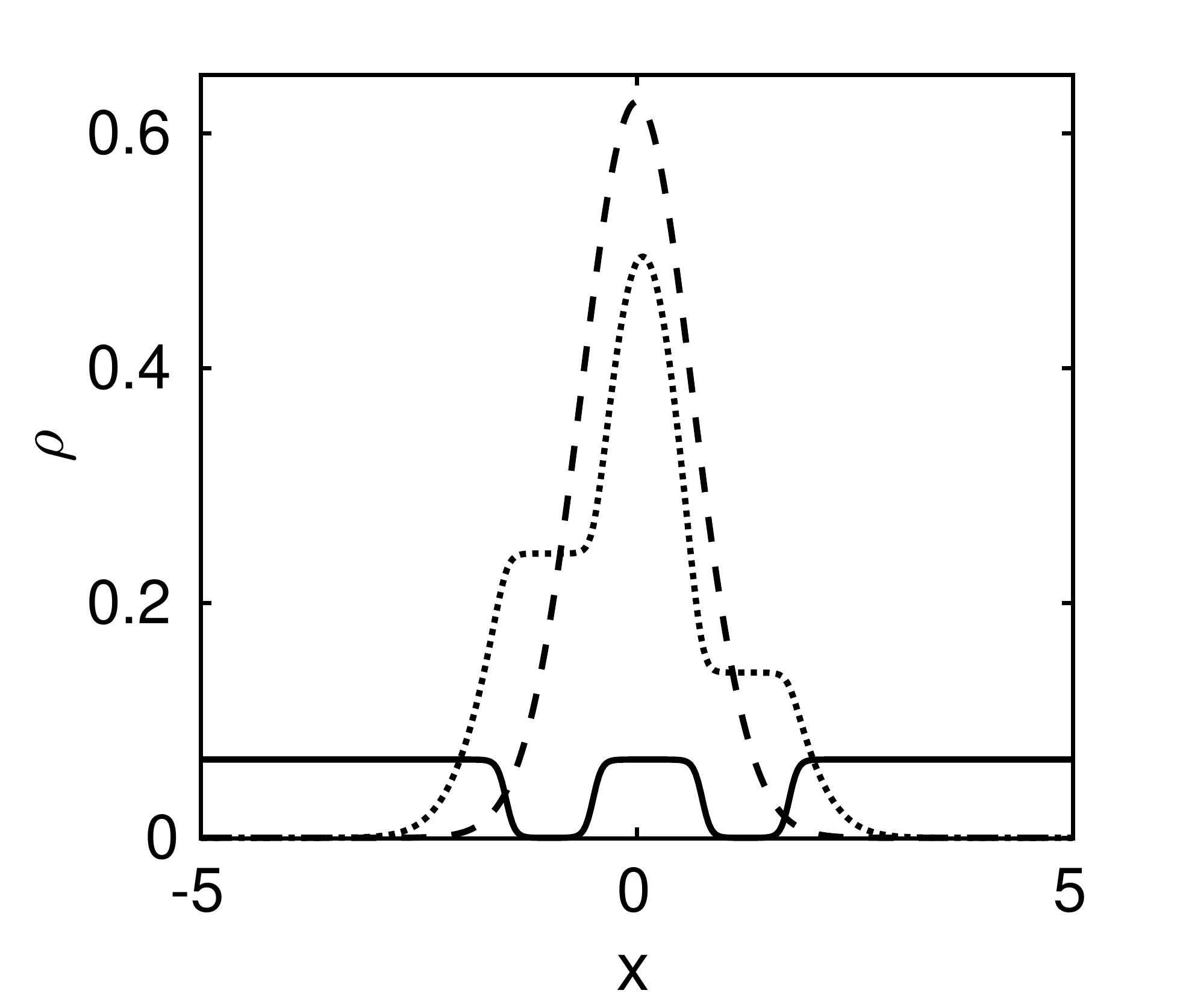}} \\b)
	\end{minipage}
	\vfill
	\begin{minipage}[h]{0.49\linewidth}
		\center{\includegraphics[width=1\linewidth]{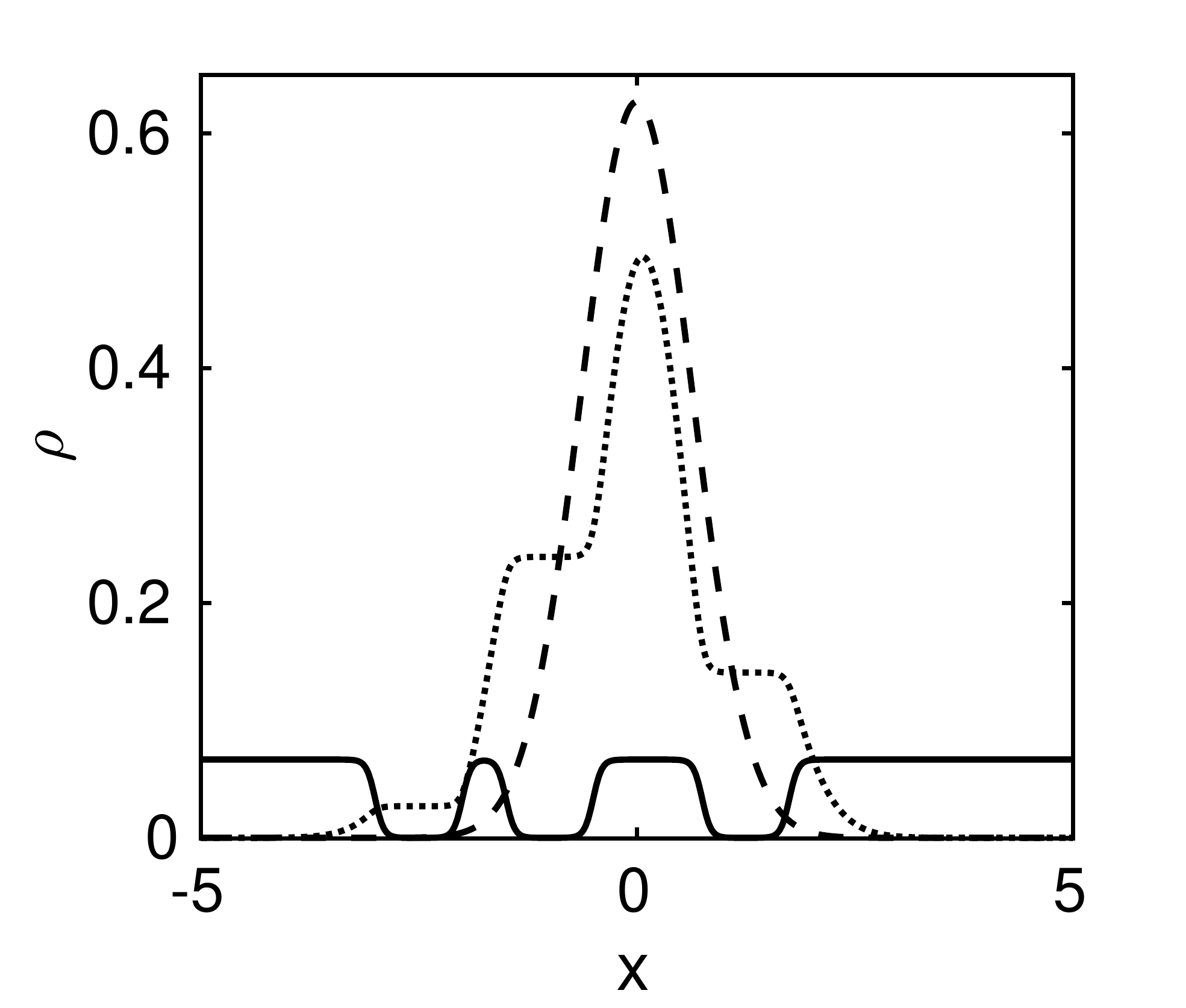}} c) \\
	\end{minipage}
	\hfill
	\begin{minipage}[h]{0.49\linewidth}
		\center{\includegraphics[width=1\linewidth]{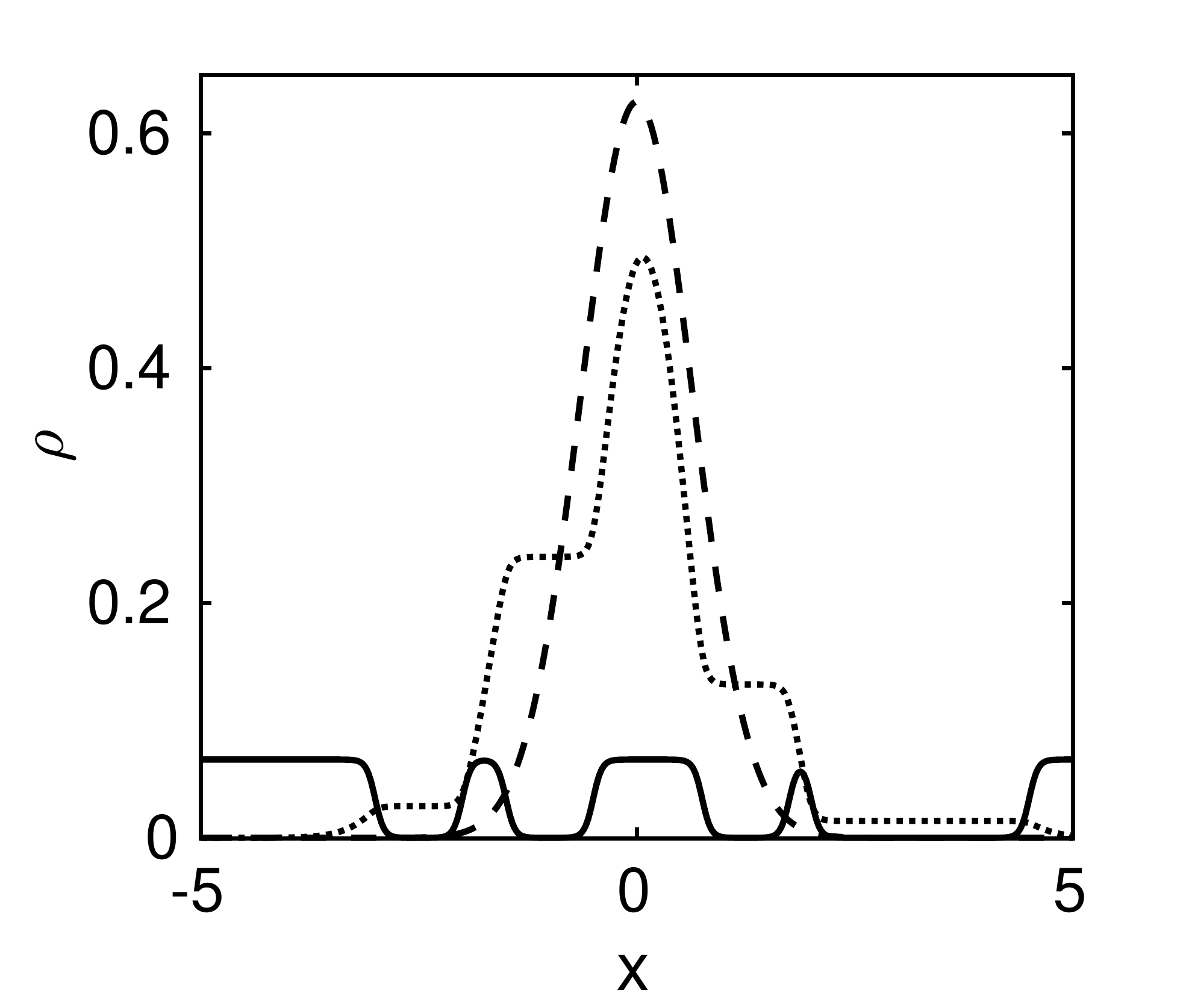}} d) \\
	\end{minipage}
	\caption{Influence of space function $\chi$ on PDF $\rho$. Gauss PDF is dashed line, PDF from equation (\ref{FPresult}) -- dotted line, space function $\chi/15$ represented with solid line. a) one tube, b)
		two CNTs, c) three CNTs, d) four CNTs.}
	\label{fig:pdfs}
\end{figure}

To solve (\ref{FPresult}) we moved on to dimensionless quantities: $t\rightarrow~\tau t$, $x \rightarrow l x$, $m \rightarrow m/l$, $D \rightarrow D l^2/\tau$; where $l$ and $\tau$ are characteristic parameters of length and time. 
Now, physical values of interest may be obtained by appropriate choice of $l$ and $\tau$.
Thereafter, numerical solutions of dimensionless equation (\ref{FPresult}) were obtained and resulting PDFs together with Gaussian PDF and $\chi$ space function are presented in Fig. \ref{fig:pdfs} for cases of one, two, three and four CNTs. For convenience on figure $\chi$ was reduced by $15$ times.

\begin{table}
	\caption{\label{tab:table1}
		Parameters of four models, presented in Fig. \ref{fig:pdfs}:  first column - name of a model, responding to models in Fig. \ref{fig:pdfs}; second column - coordinates of CNT; third, fourth columns - parameters, obtained from approximation of mean square displacement: $\langle x^2 \rangle \simeq 2At^{\mu}$.
	}
	\begin{center}
		\begin{tabular}{cccc}
			\hline\hline
			\textrm{Model}&
			\textrm{Coordinates}&
			\textrm{$A$}&
			\textrm{$\mu$}\\
			\hline
			a & $x_0=0.75, x_1=1.75$ & $0.02066$ & $1.168$ \\
			& & & \\
			& $x_0=0.75, x_1=1.75$ & \vspace{-0.2cm} \\
			\vspace{-0.2cm} b & & $0.04406$& $1.069$ \\
			& $x_{-1}=-0.5, x_{-2}=-1.5$  &  & \\
			& & &\\          
			& $x_0=0.75, x_1=1.75$ & & \\
			c & $x_{-1}=-0.5, x_{-2}=-1.5$ & $0.03891$ & $1.137$ \\
			& $x_{-3}=-2, x_{-4}=-3$ & & \\
			& & &\\          
			& $x_0=0.75, x_1=1.75$ & \\
			& $x_{-1}=-0.5, x_{-2}=-1.5$  & \vspace{-0.2cm}\\
			\vspace{-0.2cm} d & & $0.02656$ & $1.343$ \\
			& $x_{-3}=-2, x_{-4}=-3$ & \\
			& $x_2=2, x_3=4.5$ & & \\ 
			\hline\hline
		\end{tabular}
	\end{center}
\end{table}

For our model the following parameters were chosen. Boundary conditions are $\rho(\infty,t)=\rho(-\infty,t)=0$, initial condition is $\rho(x,0)=\delta(x)$, time of calculation is $t=20$ and diffusion coefficient is $D=10^{-2}$. Degree of approximation of $\chi$ function (\ref{chiDef}) is $m=10$. Coordinates of CNTs are given in Table \ref{tab:table1}.

As it can be seen from Fig.~\ref{fig:pdfs} PDFs, obtained from equation (\ref{FPresult}), do not change for areas, where CNTs are located (excepting the boundaries). This effect takes place for one, two, three, four CNTs. Thus it is shown that for final PDF different results can be reached. This is connected not only with number of CNTs, but also with their size and spatial arrangement relatively $x=0$.

\begin{figure}[h]
	\begin{minipage}[h]{0.49\linewidth}
		\center{\includegraphics[width=1\linewidth]{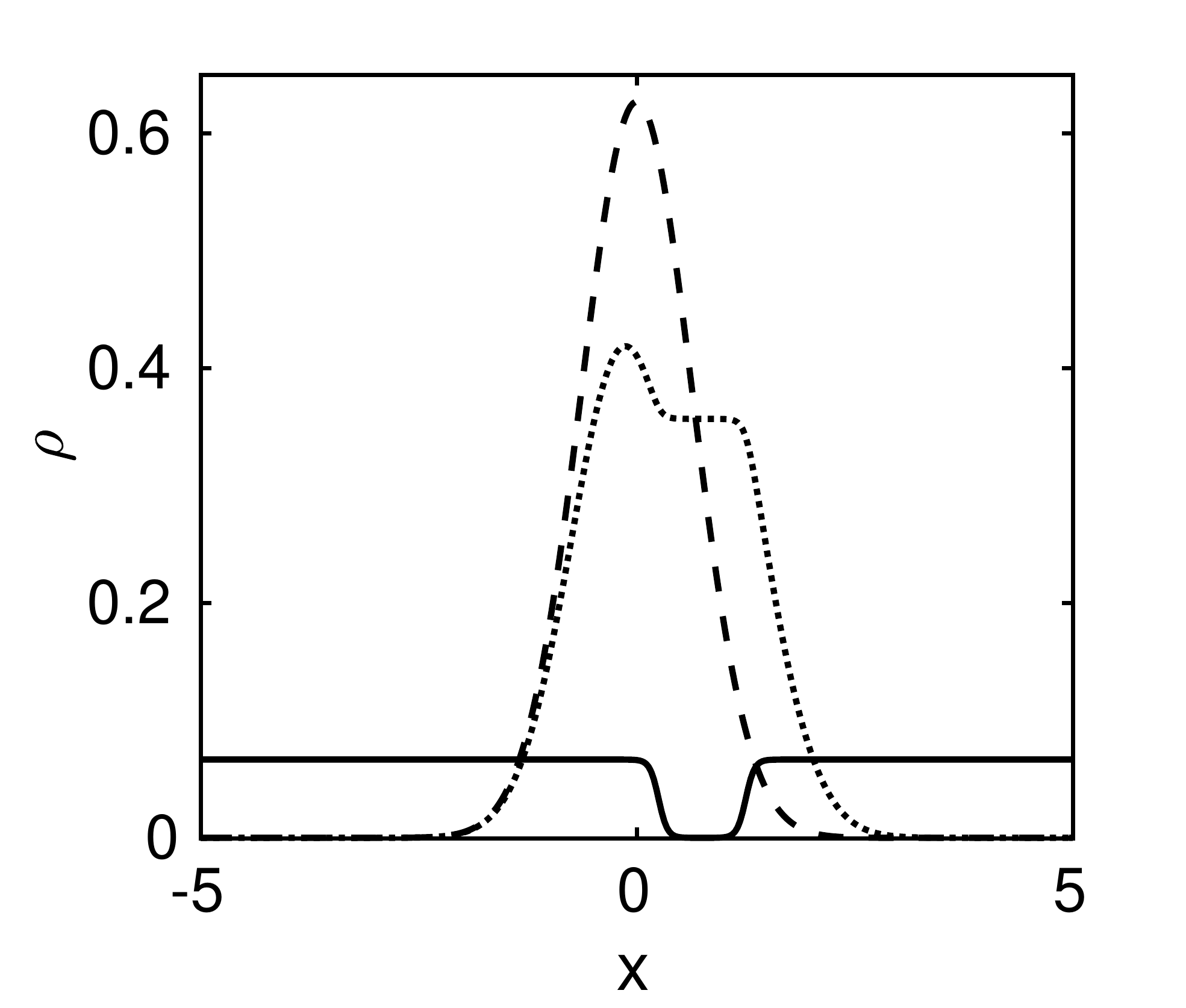}} a) \\
	\end{minipage}
	\hfill
	\begin{minipage}[h]{0.49\linewidth}
		\center{\includegraphics[width=1\linewidth]{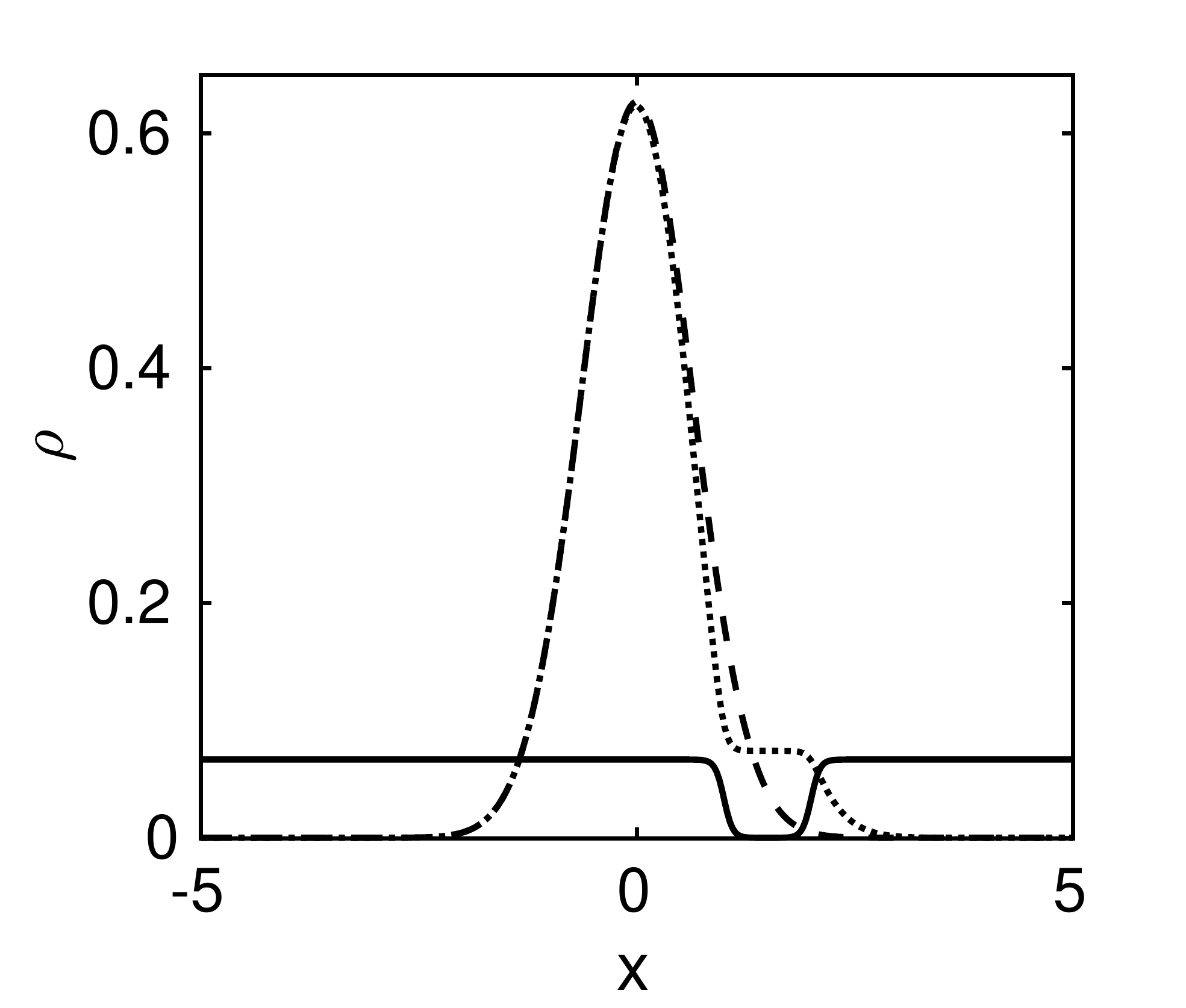}} \\b)
	\end{minipage}
	\caption{Effect of CNT displacement. Coordinates of CNT are a) $x_0=0.25, x_1=1.25$ b) $x_0=1, x_1=2$. The notation is the same as in Fig. \ref{fig:pdfs}.}
	\label{fig:pdfs1}
\end{figure}

We examine the following situation to find out how the position of the tube relative to zero affects on the PDF. Let's fix a size of a tube, and then change its position. Clearly, if CNT is located near to $x=0$, probability that particle reaches it area for time $t$ is more than probability of reaching of remote CNT. Hence first distribution is more different from the Gaussian PDF than the latter (Fig. \ref{fig:pdfs1}).

\begin{figure}[h]
	\begin{minipage}[h]{0.49\linewidth}
		\center{\includegraphics[width=1\linewidth]{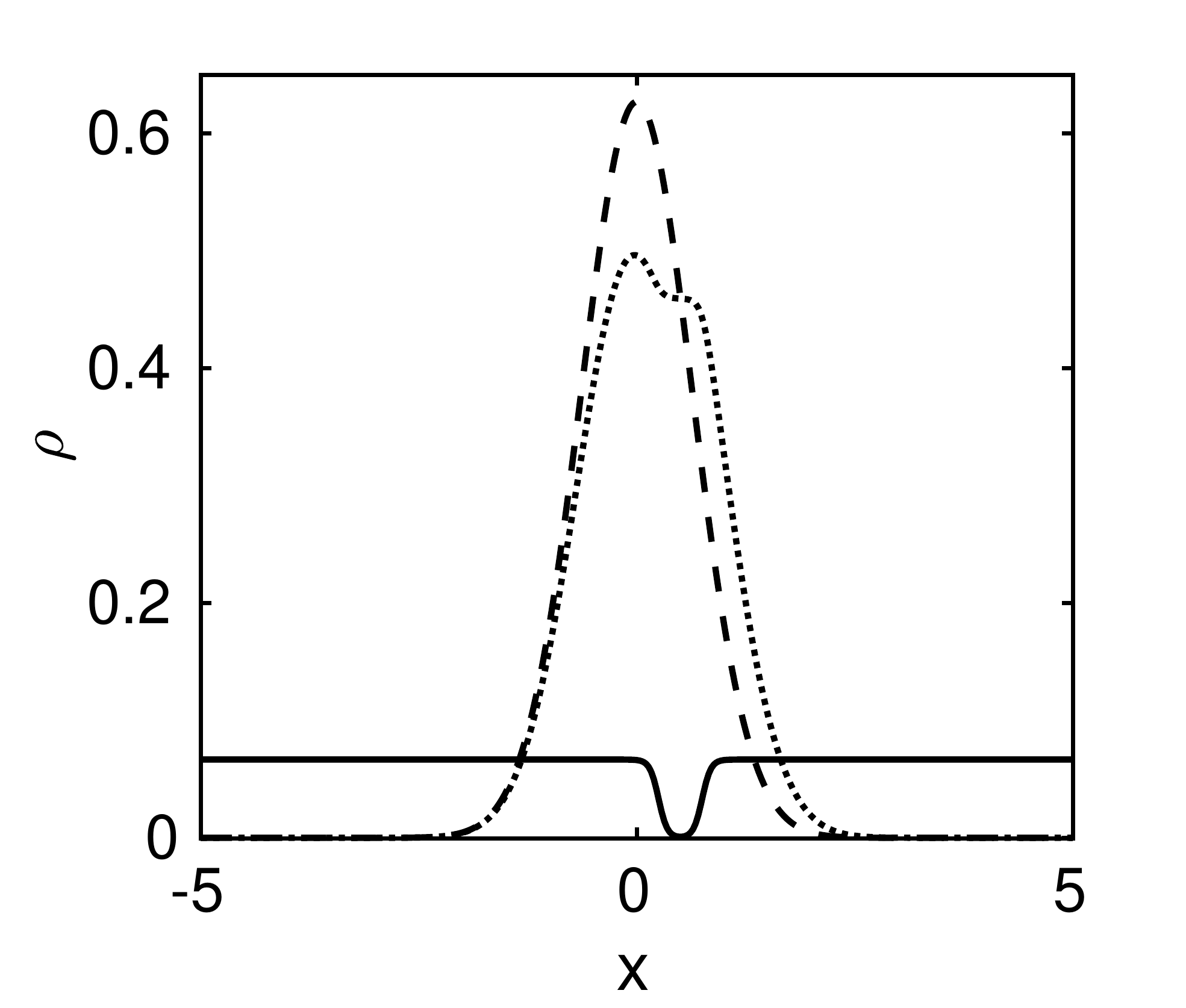}} a) \\
	\end{minipage}
	\hfill
	\begin{minipage}[h]{0.49\linewidth}
		\center{\includegraphics[width=1\linewidth]{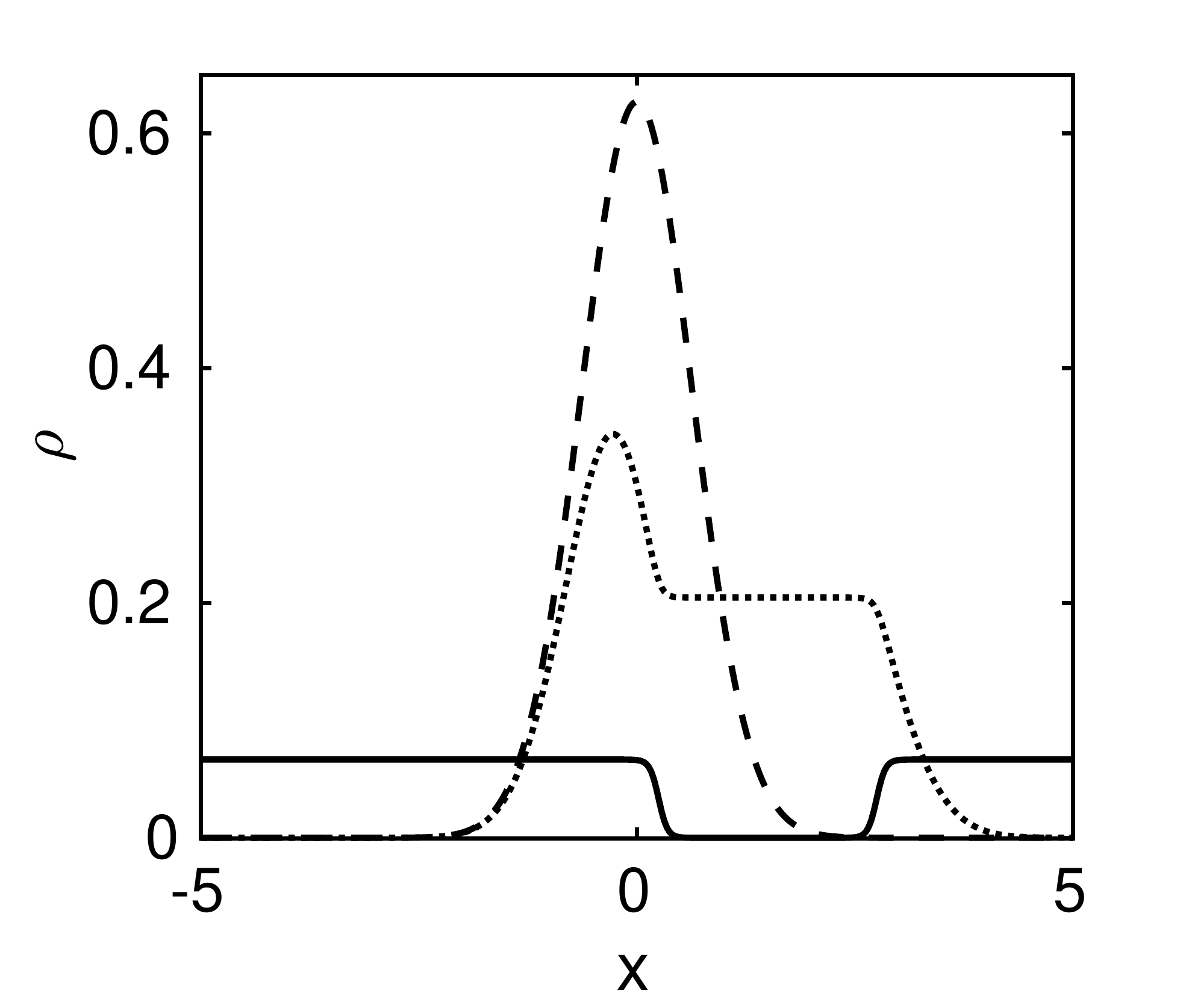}} \\b)
	\end{minipage}
	\caption{Effect of CNT length. Coordinates of CNT are a) $x_0=0.25, x_1=0.75$ b) $x_0=0.25, x_1=2.75$. The notation is the same as in Fig. \ref{fig:pdfs}.}
	\label{fig:pdfs2}
\end{figure}

For different tube sizes it is logical to assume the following: for a longer tube difference between gotten PDF and Gaussian PDF should be greater. It is demonstrated in Fig. \ref{fig:pdfs2}. Thus the competition of three parameters (numbers of CNTs, their size and spatial arrangement relatively $x=0$) is decisive for determining the final transfer regime.

\begin{figure}[h]
	\begin{minipage}[h]{0.9\linewidth}
		\center{\includegraphics[width=1\linewidth]{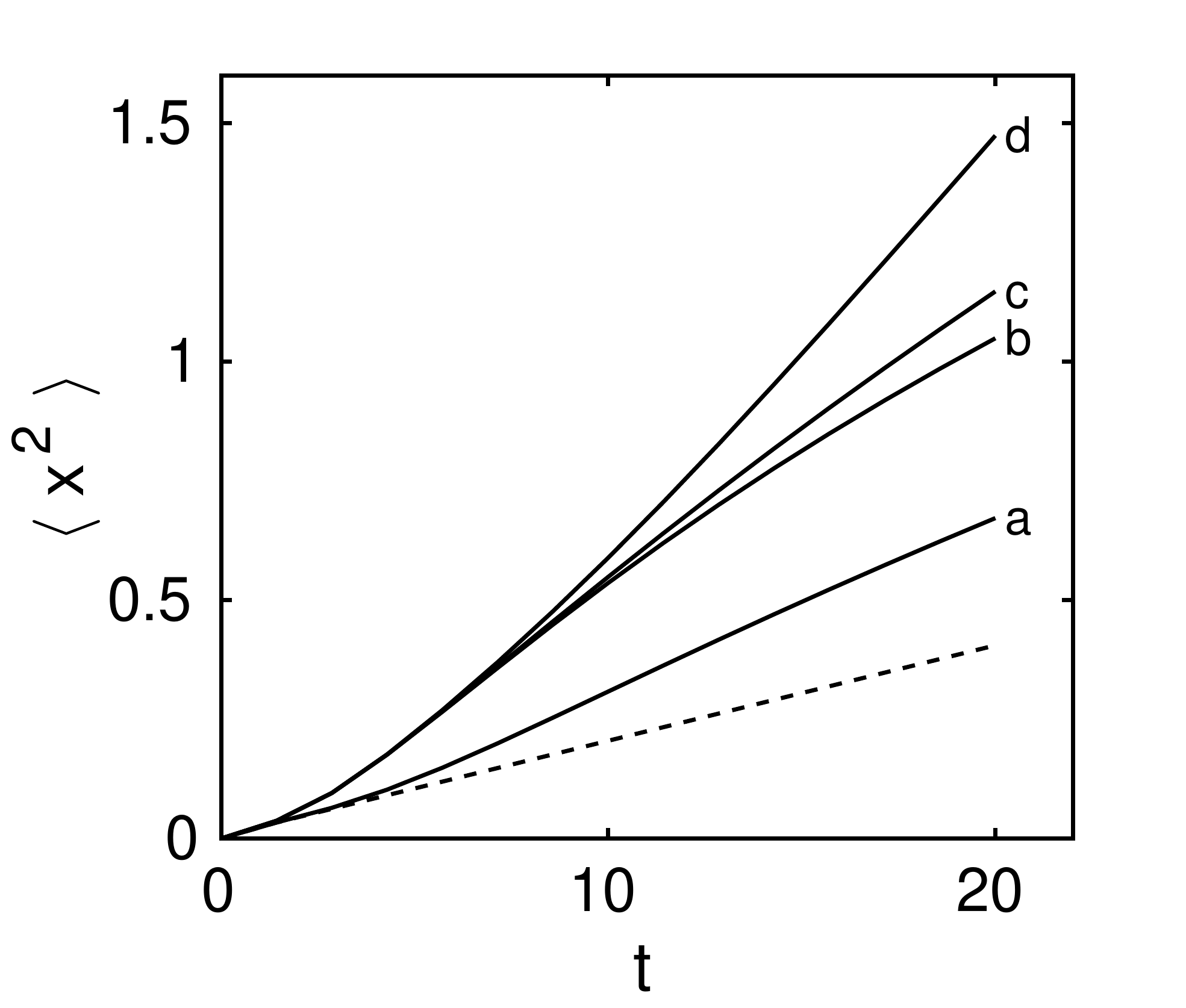}}  \\
	\end{minipage}
	\caption{Mean square displacement as function of time. Solid lines: a)~One tube, b)
		two CNTs, c) three CNTs, d) four CNTs; dashed line corresponds to Gauss distribution.}
	\label{fig:MSD}
\end{figure}

In Fig. \ref{fig:pdfs} it is shown that increasing of number of CNTs leads to increasing of deviation from Gauss PDF. Obviously, such deviation should change the MSD - it should not obey the linear time dependence. Our calculations confirm this, which can be seen in Fig. \ref{fig:MSD}. Then we fitted MSD as power law $\langle x^2 \rangle = 2At^{\mu}$, obtained $A$ and $\mu$ are presented in Table \ref{tab:table1}. However, as noted above, CNTs spatial configuration also affects on the final PDF. As a result, exponent $\mu$ depends nonmonotonically on number of tubes.

\begin{figure}[h]
	\begin{minipage}[h]{0.9\linewidth}
		\center{\includegraphics[width=1\linewidth]{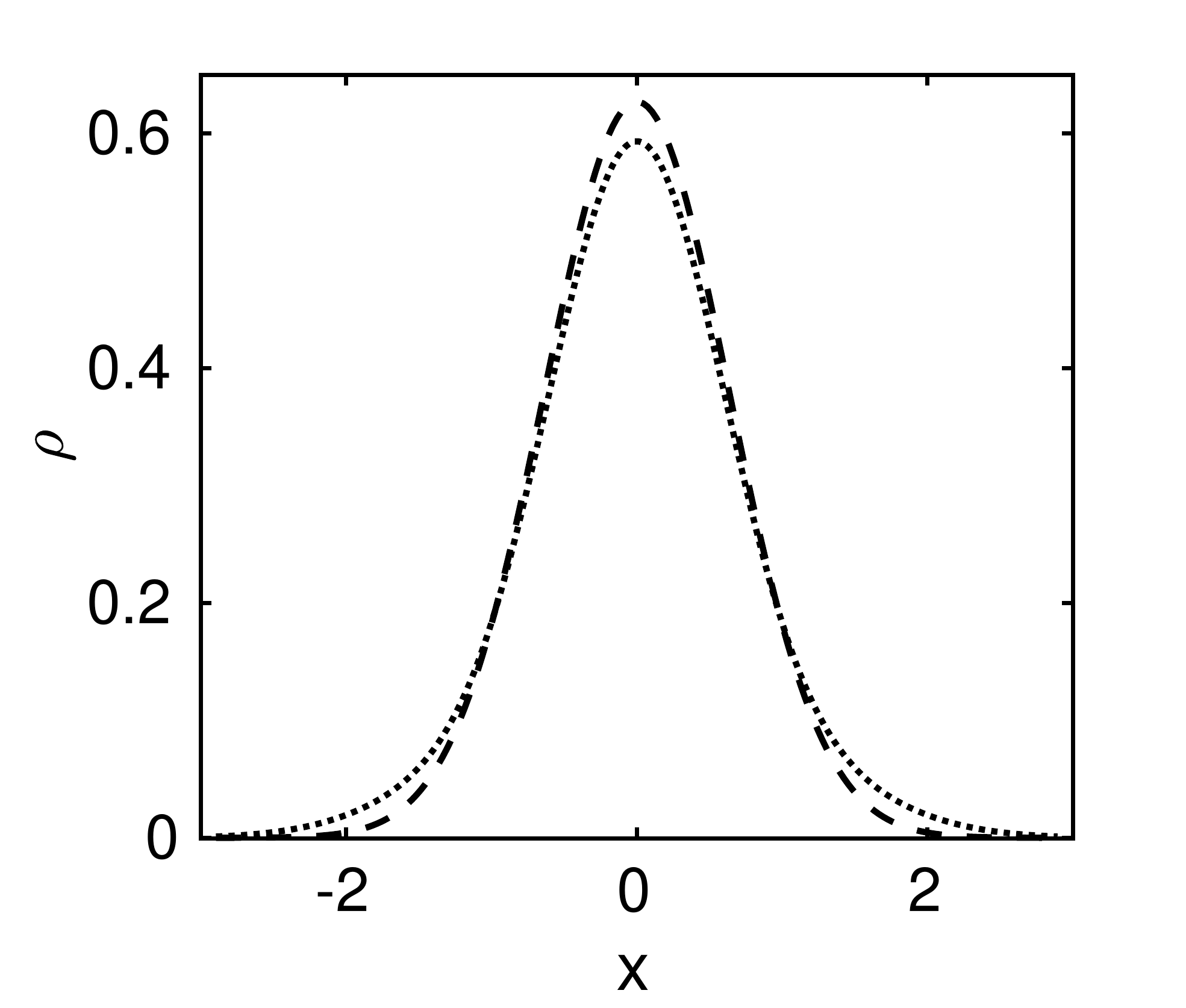}}  \\
	\end{minipage}
	\caption{PDFs of equation (\ref{FPresult}), averaging on spatial configuration for one CNT fixed length - dotted, Gauss PDF - dashed.}
	\label{fig:smooth3}
\end{figure}

Usually we cannot know the exact positions for CNTs from experimental data,
however, it is known that in experiments they try to achieve a uniform distribution of the filler in the polymer. Therefore we can average PDFs over several different configurations for CNTs, which have uniform spatial distribution, as in our previous work. In this case, final PDF should be smoothed out and will no longer be a constant function on the some regions. This PDF is presented on Fig. \ref{fig:smooth3} together with Gaussian PDF. Approximation its distribution as stable gave characteristic exponent $\alpha=1.845$, whereas approximation for Gauss distribution gave $\alpha=2$. Thus we got a result that is qualitatively consistent with the result of our previous work \cite{Likhomanova_2018}.

\section{Conclusion}
We investigated diffusion of gases for heterogeneous environment, which qualitatively reflects the transport properties of a polymeric membrane with CNTs. It was supposed that transfer regime of gaseous particle is diffusion for polymer and ballistic for CNT regions. Under this assumption Fokker-Planck equation was derived from stochastic equations. There was sequentially shown how solution of Fokker-Planck equation is changed in presence of CNT areas. Influence of numbers of CNTs, their size and spatial arrangement relatively $x=0$ was analyzed. The obtained results generally demonstrate that increasing of CNT numbers leads to increasing of exponent $\mu$ for mean square displacement in expression $\langle x^2 \rangle \sim t^{\mu}$. Additionally, this exponent is also governed by spatial distribution of CNTs relatively $x=0$ and CNTs length. 
Presented results show nonlinear time dependence of mean square displacement, that, as rule, is connected with anomalous diffusion regime. Also, in this work stable distribution as result of averaging on different spatial configuration of CNTs of fixed length was obtained. Thus, appearance of anomalous diffusion regime was explained in frames of linear Fokker-Planck equation without using apparatus of fractional differential equation. Presented results qualitatively confirm to results of \cite{Likhomanova_2018}.

\section*{Acknowledgement}

The second author's work was supported by the Research Center “Kurchatov Institute” (order No.~1878~of~08/22/2019).

\end{document}